\documentclass[journal]{vgtc}                     

\onlineid{1699} 

\vgtccategory{Theoretical \& Empirical}

\title{How Data Journalists Visualize Uncertainty: Editorial Constraints and Design Heuristics in Newsroom Practice}

\author{
\authororcid{Carmen Garro}{0009-0007-7744-0994},
  Annemarie Wisbar,  
  Benjamin Possin, and
  \authororcid{Christoph Kinkeldey}{0000-0001-5669-6295}
}

\authorfooter{
  \item
  	Carmen Garro is with viu:lab at the Hamburg University of Applied Sciences.
  	E-mail: Carmen.GarroCarranza@haw-hamburg.de.
  \item
  	Annemarie Wisbar is with viu:lab at the Hamburg University of Applied Sciences.
  	E-mail: Annemarie.Wisbar@haw-hamburg.de.
  \item
  	Benjamin Possin is with viu:lab at the Hamburg University of Applied Sciences.
  	E-mail: Benjamin.Possin@haw-hamburg.de
  \item
  	Christoph Kinkeldey is the head of viu:lab at the Hamburg University of Applied Sciences.
  	E-mail: Christoph.Kinkeldey@haw-hamburg.de.
}

\abstract{%
  Communicating uncertainty is increasingly recognized as important in data journalism, yet in practice it is often oversimplified, downplayed, or omitted. 
  To understand how uncertainty is handled in newsroom visualization practice, we conducted a full-day workshop with 11 data journalists from major German-speaking newsrooms. The workshop combined presentations of newsroom examples, collaborative scenario classification, World Café discussions on authoring hurdles, and group reflections on audience needs and future directions. 
  Based on a thematic analysis of workshop materials and transcripts, we examine how journalists currently approach uncertainty visualization in newsrooms. We find that uncertainty communication is treated as a workflow- and context-dependent editorial decision rather than primarily as a visualization problem. It is further constrained by production overhead, anticipated audience misinterpretation, publication formats, and the need to preserve a clear takeaway. Rather than relying on advanced visualization techniques, practitioners favor pragmatic defaults, layered communication, and explanatory guidance tailored to audience and medium. We also find that the scenario-based framework we provided is valued not as a strict taxonomy for classifying finished visualizations, but as a planning prompt for early design reasoning. 
  Based on these insights, we contribute (i) practitioner-informed findings on the challenges and strategies of visualizing uncertainty in journalistic practice, (ii) a refined scenario-based framework to support communication about uncertainty, and (iii) practice-oriented heuristics for designing uncertainty visualizations in newsrooms. 
  Our work contributes to bridging visualization research and newsroom practice by grounding uncertainty communication in the realities of journalistic production.
}


\teaser{
  \centering
  \includegraphics[width=\linewidth, alt={A view of a city with buildings peeking out of the clouds.}]{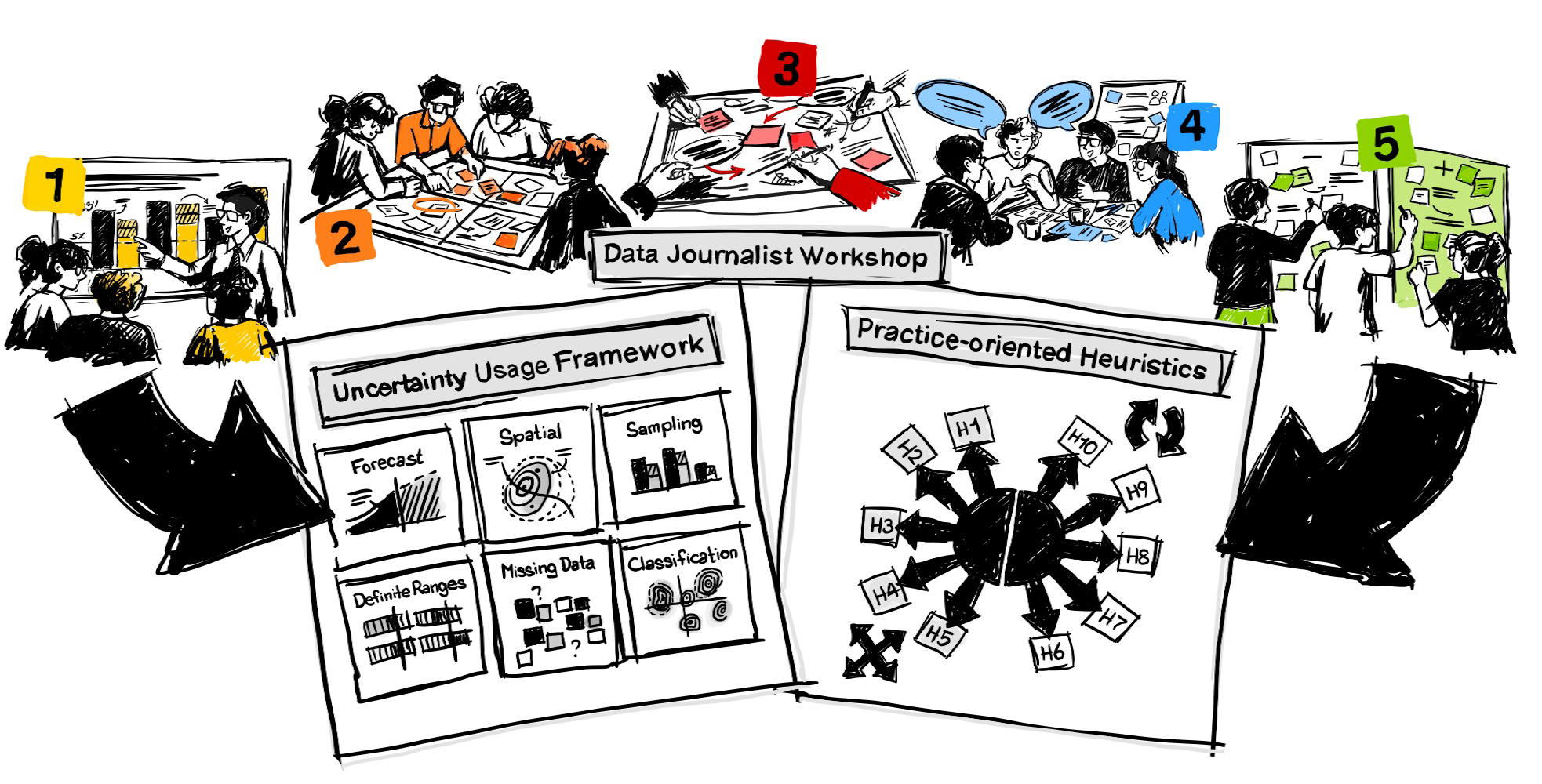}
  
  \caption{
Overview of our workshop with 11 data journalists about current uncertainty-visualization practice, challenges, and needs in newsroom production. The workshop consisted of five phases: \roundnum{color_phase1}{1} Newsroom examples, \roundnum{color_phase2}{2} UUS framework classification, \roundnum{color_phase3}{3} Authoring hurdles, \roundnum{color_phase4}{4} Audiences, and \roundnum{color_phase5}{5} Takeaways \& future directions.
Based on the workshop, we provide practitioner-informed findings, a refined
scenario-based UUS framework,
and practice-oriented heuristics. 
  }
  \label{fig:teaser}
}

\graphicspath{{figs/}{figures/}{pictures/}{images/}{./}} 

\usepackage[dvipsnames]{xcolor}
\usepackage[normalem]{ulem}

\definecolor{christoph}{rgb}{0.9, 0.5, 0.1}

\newcommand{\cg}[1]{\textcolor{RoyalBlue}{#1}}
\renewcommand{\cg}[1]{#1}
\renewcommand{\sout}[1]{}

\usepackage{tikz}
\usepackage{xcolor}
\definecolor{color_phase1}{HTML}{fff4b9}
\definecolor{color_phase2}{HTML}{ffdcbe}
\definecolor{color_phase3}{HTML}{f4bfbf}
\definecolor{color_phase4}{HTML}{bee1f2}
\definecolor{color_phase5}{HTML}{ebf4bc}

\usepackage{tikz,xcolor}

\newcommand{\circnum}[2]{
  \tikz[baseline=(n.base)]{
    \node[circle, draw=#1, fill=white, line width=0.8pt,
          inner sep=0.8pt] (n) {\textcolor{#1}{\bfseries #2}};
  }%
}

\newcommand{\roundnum}[2]{
  \tikz[baseline=(n.base)]{
    \node[draw=#1, fill=#1, rounded corners=2pt, line width=0.8pt,
          inner xsep=3pt, inner ysep=1.2pt] (n)
          {\textcolor{black}{\bfseries #2}};
  }%
}

\usepackage{etoolbox}
\robustify{\roundnum}

\usepackage{multirow}
\usepackage{float}

\usepackage{tabu}                      
\usepackage{booktabs}                  
\usepackage{lipsum}                    
\usepackage{mwe}                       
\usepackage{ccicons}                   

\usepackage{graphicx}
\usepackage{tabularx}
\usepackage{array}
\usepackage{pdflscape}

\usepackage{placeins}

\usepackage{longtable}
\usepackage{rotating}

\newcolumntype{Y}{>{\raggedright\arraybackslash}X}
\newcolumntype{C}[1]{>{\centering\arraybackslash}m{#1}}

\usepackage{mathptmx}                  

\begin{document}



\maketitle

\thanks{\textit{This work has been submitted to the IEEE for possible publication. Copyright may be transferred without notice, after which this version may no longer be accessible.}}

\section{Introduction} 

Communicating uncertainty in data journalism (\textit{a form of journalism that uses data as both a source and a tool to investigate, explain, and communicate stories}) is essential for transparency, ethical responsibility, and informed interpretation by the audience. Accordingly, it is increasingly acknowledged as a critical component of data-driven journalistic work~\cite{gray_data_2019} and has become established in specific applications such as election poll reporting~\cite{Yang:2023} and climate projections \cite{Kause:2021}. Insights from related fields suggest that in practical settings, uncertainty is often minimized or excluded altogether~\cite{Hullman:2019}, and data journalism faces additional constraints: fast-paced editorial workflows often leave little time or capacity to accommodate the additional effort required for computing, designing, and explaining uncertainty displays. Empirical studies suggest that in certain situations, communicating uncertainty can reduce trust, increase confusion, or discourage audience engagement~\cite{vanderbles2020effects}. This makes its use particularly challenging in journalistic practice. By engaging with these challenges, this research aims to identify the conditions under which uncertainty visualization can effectively support journalistic practice under production constraints.

\begin{figure*}
    \centering
    \includegraphics[width=\textwidth]{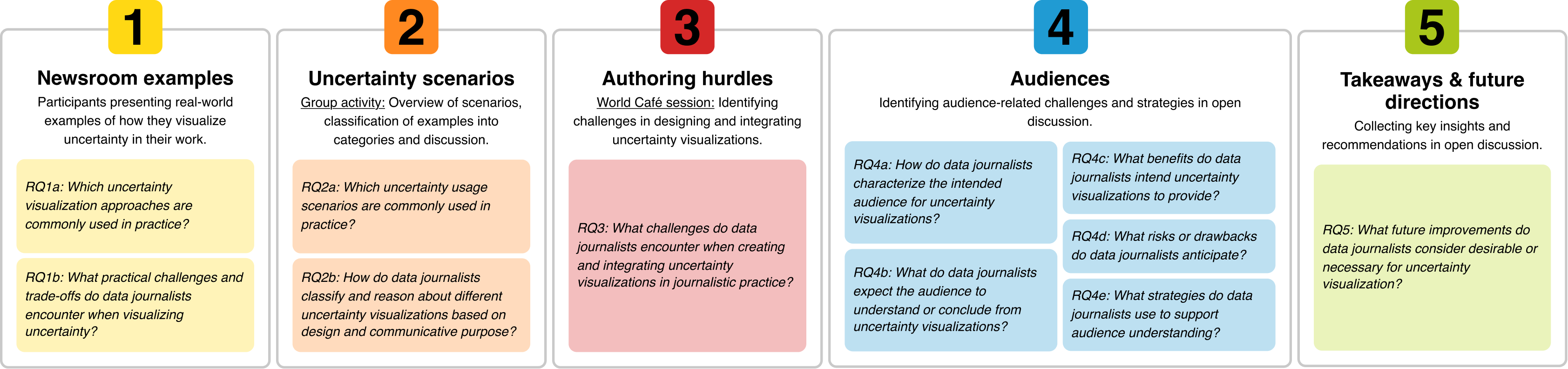}
    \caption{Five phases of the workshop, linking activities to research questions on uncertainty visualization practices, challenges, audiences, and future directions.}
    \label{fig:workshop-structure}
\end{figure*}

Our guiding research question is \textit{``How do data journalists communicate uncertainty under newsroom constraints and audience considerations, and what does this imply for uncertainty visualization design?''}. Supporting research questions (Figure~\ref{fig:workshop-structure}) address key aspects of the use of uncertainty visualizations in data journalism, including current practices, application scenarios, authoring challenges, audience perspectives, and future needs. To investigate this question we conducted a full-day \textit{workshop with data journalists} from major German-speaking newsrooms. 
Based on our research questions, we structured the workshop into an introduction and five phases aimed at encouraging both reflection and interaction among participants: sharing newsroom examples, small-group scenario classification using our UUS (Uncertainty Usage Scenarios) framework 
, a World Caf\'e on authoring hurdles, a plenary discussion on audience considerations, and a final synthesis using a Wall of Thoughts exercise and group discussion.

In this paper, we contribute 
(i) practitioner-informed insights into challenges and strategies for uncertainty visualization in journalistic contexts, (ii) a refined scenario-based framework for uncertainty visualization in data journalism, and (iii) practice-oriented \cg{, research-backed} heuristics to support uncertainty visualization decisions in newsrooms. Together, these contributions help narrow the gap between uncertainty visualization research and newsroom practice.
\section{Background}

Despite growing awareness of the role of uncertainty in journalistic reporting, its integration into journalistic visuals remains inconsistent. Visualization practitioners, including journalists recognize its value, but often choose to omit it. This happens mainly due to concerns that it can reduce audience comprehension or undermine the credibility of information~\cite{Hullman:2019}. This reflects a trade-off between narrative clarity and scientific transparency.

This section reviews the current state of uncertainty communication in journalism. It identifies three core challenges: first, common uncertainty encodings are prone to misinterpretation, even among experts; second, newsroom constraints pose specific challenges to the process of uncertainty visualization design; and third, existing design guidance is often not designed for use in practice.

\cg{While uncertainty visualization has been studied extensively across domains such as climate science, geospatial analysis, and decision support, researchers consistently emphasize that communicating uncertainty involves balancing transparency with usability rather than simply displaying more information. Surveys and design overviews describe uncertainty visualization as an active research area with many available techniques, yet also highlight that selecting appropriate representations remains highly context-dependent \cite{Boukhelifa:2023, Mehta:2022, Hadjimichael:2024}. This challenge becomes particularly pronounced in journalism, where visualizations must support rapid comprehension by diverse audiences while fitting within editorial constraints.}

\subsection{Interpretation Challenges in Uncertainty Visualization}
Visual complexity can increase cognitive demands and reduce judgment accuracy~\cite{Chakraborty:2024,ryan2018glance}. Textual explanations and appropriately designed uncertainty representations can support reasoning and decision making~\cite{Banchilhon:2023,Fernandes:2018}. Prior work also suggests that audiences can interpret uncertainty when it is transparently communicated and contextually framed~\cite{vanDerBles2019}.

However, interpretation of uncertainty remains difficult, and even experts are not immune to misinterpreting uncertainty visualizations, e.g., when distinguishing confidence intervals from standard error bars or when accounting for the implications of study design~\cite{Belia:2005}. Known misinterpretations include ``within-the-bar'' bias, where error bars are often mistaken for uniform probability regions~\cite{Hofman:2020} and the ``deterministic construal error'', where viewers interpret visualizations deterministically, ignoring the uncertainty depicted in the graphic~\cite{Joslyn:2021}. Alternatively, if uncertainty is not shown, there is no guarantee that readers will infer the distribution correctly~\cite{Hullman:2019}, leaving space for a ``free-for-all'' interpretation of the underlying distribution.

These interpretation challenges contribute to the omission of uncertainty in practice, along with concerns about confusing or overwhelming readers, fear of making data seem questionable, the belief that omitting uncertainty will simplify decision making, and a concern that it undermines credibility~\cite{Hullman:2019}.

\cg{More generally, researchers have argued that uncertainty should not be understood solely as a visualization challenge but as a communication challenge. Effective uncertainty communication depends not only on graphical encodings but also on contextual explanations, audience expectations, and the purpose of the visualization \cite{Marcot:2020, Fischhoff:2012}.}



\subsection{Constraints of Uncertainty Communication in Newsroom Practice} 

Designing uncertainty visualizations in data journalism is constrained by time pressure, limited resources, and limited insights into audience characteristics such as statistical literacy. These challenges become particularly visible in \textit{predictive journalism}, an emerging form of reporting where forecasts and model outputs are communicated to broad audiences. This has prompted innovations in narrative formats, interactive tools, and visual design to communicate uncertainty responsibly~\cite{Pentzold:2021, Pentzold:2020}.

In addition, information about the uncertainty is often not readily available or straightforward to derive in journalistic production, leaving practitioners without clear measures to compute and visualize (e.g., \cite{Hullman:2019,Morini:2023}).
A further constraint is audience heterogeneity. Research has not yet told us much about how to address a public with widely varying statistical and visual literacy, which is precisely the reality of journalism. \cg{Similar challenges have been identified across scientific uncertainty communication, where audience characteristics such as prior beliefs, numeracy, and contextual knowledge substantially influence how uncertainty is interpreted. Consequently, there is growing recognition that uncertainty communication strategies cannot be universally effective across all audiences \cite{Bryan:2021, Kause:2021}.} The field of data journalism still has the challenge of which uncertainty representations are effective for which audiences and how to develop them under newsroom constraints. 
Evidence on audience effects is mixed. For example, representing numerical ranges with point estimates does not decrease trust, whereas verbal uncertainty can slightly reduce perceived reliability ~\cite{vanderbles2020effects}. Audiences can engage with uncertainty when explanations or interactive tools clarify its implications. Moreover, audience characteristics such as numeracy, education, prior beliefs, and trust, mediate the effectiveness of uncertainty communication~\cite{Yang:2024, Yang:2023, krause24, Weiskopf:2022, Allaham:2022}.

\subsection{Design Guidance for Newsroom Practice}\label{subsec:guidance} 
Despite extensive research on uncertainty communication, practice-oriented guidance tailored to newsroom work remains limited. Existing studies and frameworks explain why uncertainty is often omitted but provide less insight into how journalists and other practitioners navigate trade-offs between transparency and clarity under real production constraints. More broadly, research has documented challenges when transferring visualization guidance into practice, including in uncertainty visualization~\cite{Boukhelifa:2023,Mehta:2022,Jena:2020}.

A key reason is that many guidelines implicitly assume time for explanation and development of the visualization and a level of statistical literacy that are difficult to rely on in general news contexts~\cite{Hullman:2019}. Under newsroom conditions that prioritize speed, clear takeaways, and broad comprehension~\cite{Arafat:2023,Ivask:2023}, adding uncertainty layers to a data visualization can increase cognitive load and may reduce perceived credibility if not carefully framed~\cite{Joslyn:2021,Chakraborty:2024}. This creates tensions with journalistic goals of clarity, trust, and authority~\cite{Skovsgaard:2024,Detzen:2024}.


In addition, uncertainty communication depends on production infrastructure. It often requires bridging scientific results and communication practices, sometimes involving collaboration with domain experts~\cite{Marcot:2020}. Integration of uncertainty-aware data pipelines into editorial workflows poses further technical and logistical hurdles~\cite{Morini:2023}, especially in organizations that are still building data journalism capacity~\cite{Morini:2023}. Accordingly, interactive and scenario-based uncertainty presentations are mainly reported as case-based exemplars rather than routine newsroom output (e.g.,~\cite{Buffett:2020}).

Taken together, these gaps motivate practitioner-centered research. Prior work suggests that uncertainty is often simplified or omitted, but less is known about \textit{how} journalists develop workable solutions and negotiate trade-offs in day-to-day production. This motivated our workshop-based investigation of uncertainty visualization practices in newsrooms. \cg{Our practitioner-centered approach also responds to broader discussions within visualization research that call for design knowledge grounded in real-world practice rather than isolated evaluation of visualization techniques. Engaging practitioners directly allows design recommendations to emerge from authentic production contexts and complements existing technique-oriented research \cite{Correll:2022}.}



\section{Workshop Design and Analysis}
We chose a workshop format with data journalism practitioners to elicit knowledge about \textit{who} communicates uncertainty in newsrooms, \textit{what} is communicated, and \textit{how} it is explained and visualized in practice. The workshop offered a structured, yet flexible space to surface knowledge and exchange between research and practice, moving beyond theoretical frameworks to ground our understanding in the practical challenges and innovative solutions developed in the field.
In May 2025, we held a one-day workshop at 
the Hamburg University of Applied Sciences
(Figure \ref{fig:workshop-day}), structured around research questions regarding uncertainty visualization practices, authoring hurdles, audience considerations, and future needs \cg{(Figure \ref{fig:workshop-structure})}.

\subsection{Participants and recruitment method}
The workshop targeted data journalists with prior experience visualizing uncertainty in reporting. We used purposive recruitment to reach practitioners with relevant expertise. Invitations were distributed through relevant professional mailing lists and shared via personal networks within data journalism communities.  In addition, we contacted selected newsrooms directly to reach data visualization teams. Participation was free and travel funds were available. Eleven participants (seven female) from seven different newsrooms attended the workshop, six based in Germany and one in Switzerland.

\cg{\subsection{Initial UUS Framework}
Prior to the workshop, we developed the Uncertainty Usage Scenarios (UUS) framework based on a structured review of uncertainty visualizations in online data journalism\cite{possin_charting_2025}. We examined approximately 900 articles published between January 2024 and April 2025 across 17 German- and English-language media outlets and identified 75 visualizations that explicitly communicated uncertainty through visual means. We documented and categorized these examples according to the uncertainty represented and their visualization characteristics, including the dimensions of Kinkeldey et al.’s \textit{uncertainty cube}\cite{kinkeldey_how_2014}. Based on this analysis, we derived recurring usage scenarios that formed the initial UUS framework. In the present study, we introduced this framework to participants in Phase~\roundnum{color_phase2}{2} to examine how these scenarios correspond to newsroom practice and how practitioners interpret and apply them to their own examples.}

\subsection{Workshop Structure}
In order to address the workshop topics, we divided the day into an introduction and five phases (Figure~\ref{fig:workshop-structure}). 

\textbf{Introduction (30 minutes).} We welcomed participants, introduced the broader research project, and outlined the purpose of the workshop.

\textbf{Phase \roundnum{color_phase1}{1}: Newsroom examples (60 minutes).} Participants were asked in advance to bring newsroom examples in which they had visualized uncertainty or experimented with uncertainty  communication. We were particularly interested in challenging cases, including unfinished or unpublished work. Participants added their examples in a shared document (screenshots, links, and brief notes) and then presented selected cases to the group.\cg{The shared document, presentation notes, and transcripts of the audio recordings served as workshop artifacts for analysis.}

\textbf{Phase \roundnum{color_phase2}{2}: Uncertainty scenarios (90 minutes).} 
\cg{We presented the scenarios defined in the UUS framework to our workshop participants. After the presentation}, participants formed small groups (2-3 people) and classified their collected examples into the scenario categories. Groups then reported back in plenary to compare classifications and discuss the relevance and limitations. Artifacts from this phase included the scenario assignments, discussion notes and the edited shared document, in addition to audio recordings of the discussion, which were transcribed and analyzed.

\textbf{Phase \roundnum{color_phase3}{3}: Authoring hurdles (60 minutes).} After lunch, participants split into groups to discuss authoring challenges in a World Caf\'e format. Four tables were prepared with four guiding questions: (i) ``In which articles / stories or situations do you decide to visualize uncertainties---and when do you consciously leave them out?'' (ii) ``Which editorial or communicative hurdles do you experience when implementing uncertainty visualizations in your contributions?'' (iii) ``Which approaches to visualizing uncertainties have not worked in the past and why?'' (iv) ``What undesirable reactions of the readership do you fear if you show uncertainty?'' Groups rotated between the tables, discussed and documented their thoughts on table sheets, and could build on the notes from prior groups (Figure \ref{fig:teaser}). We photographed the table sheets that served as artifacts for transcription and analysis.

\textbf{Phase \roundnum{color_phase4}{4}: Audiences (60 minutes).} 
This phase focused on audience needs and interpretation. It began with a short input summarizing results from a mini-study conducted by our research group on readers' comprehension of uncertainty in a real-world election-poll visualization~\cite{garro2025uncertainty}. This was followed by a discussion in which participants shared experiences, challenges, and open questions. As artifacts, we audio-recorded this discussion for transcription and analysis.

\textbf{Phase \roundnum{color_phase5}{5}: Takeaways \& future directions (30 minutes).}
In a concluding ``Wall of Thoughts'' exercise, each participant recorded a key insight or takeaway. We then held an open discussion centered on: ``What’s next?'', ``What was useful?'' and ``What was missing?''. Artifacts from this phase included photos of the Wall of Thoughts as well as notes from the final discussion which was also audio-recorded.

\begin{figure}
    \centering
    \includegraphics[width=\linewidth]{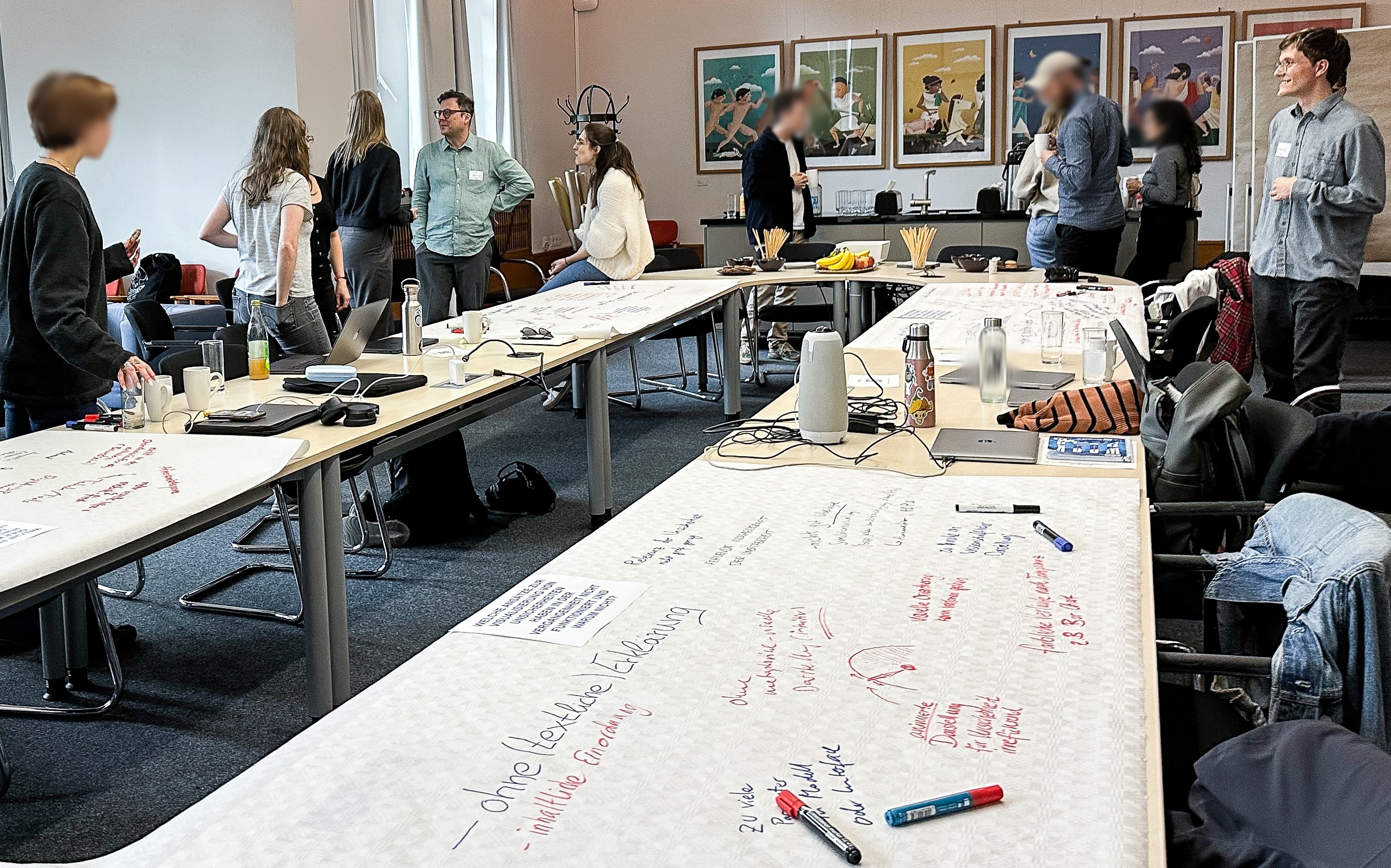}
    \caption{World Café activity during Phase \roundnum{color_phase3}{3}. Participants rotated between tables to discuss different authoring challenges, documenting their insights on shared table sheets that were later used for analysis.}
    \label{fig:workshop-day}
\end{figure}

\subsection{Data Collection and Analysis}
We collected material in all five workshop phases, including entries in a shared document and written notes produced during group activities. We additionally recorded 2.5 hours of audio from plenary discussions in Phases~\roundnum{color_phase1}{1}, \roundnum{color_phase2}{2}, \roundnum{color_phase4}{4}, and \roundnum{color_phase5}{5}, which we transcribed. 

Two team members independently conducted an inductive thematic analysis of the transcripts. We performed open coding to identify recurring topics, iteratively refined them into a shared set of codes and resolved coding differences through discussion. We segmented the transcripts into meaning units (i.e., coherent statements relevant to the discussion topic) and coded 192 segments using a code book comprising seven main categories and 20 distinct codes. 
Codes related to uncertainty scenarios were informed by our initial framework and extended with categories introduced by participants during the workshop. 
For Phase~\roundnum{color_phase3}{3}, we collected four photos of the annotated World Caf\'e table sheets (Figure~\ref{fig:workshop-day}) and two photos from the final Phase~\roundnum{color_phase5}{5}, documenting the ``Wall of Thoughts''. We identified written comments from these artifacts and mapped them to their respective research questions to cross-validate our transcript findings with the artifacts.

\section{Findings}
In the following, we present findings from the five phases of our workshop (Figure \ref{fig:workshop-structure}). We report them according to the phases, spanning newsroom examples, UUS scenarios, authoring hurdles, audience considerations, and participants' takeaways and future requirements. Each section will be identified for its reference to a workshop section with the corresponding color.


\begin{figure}
    \centering
    \includegraphics[width=\linewidth]{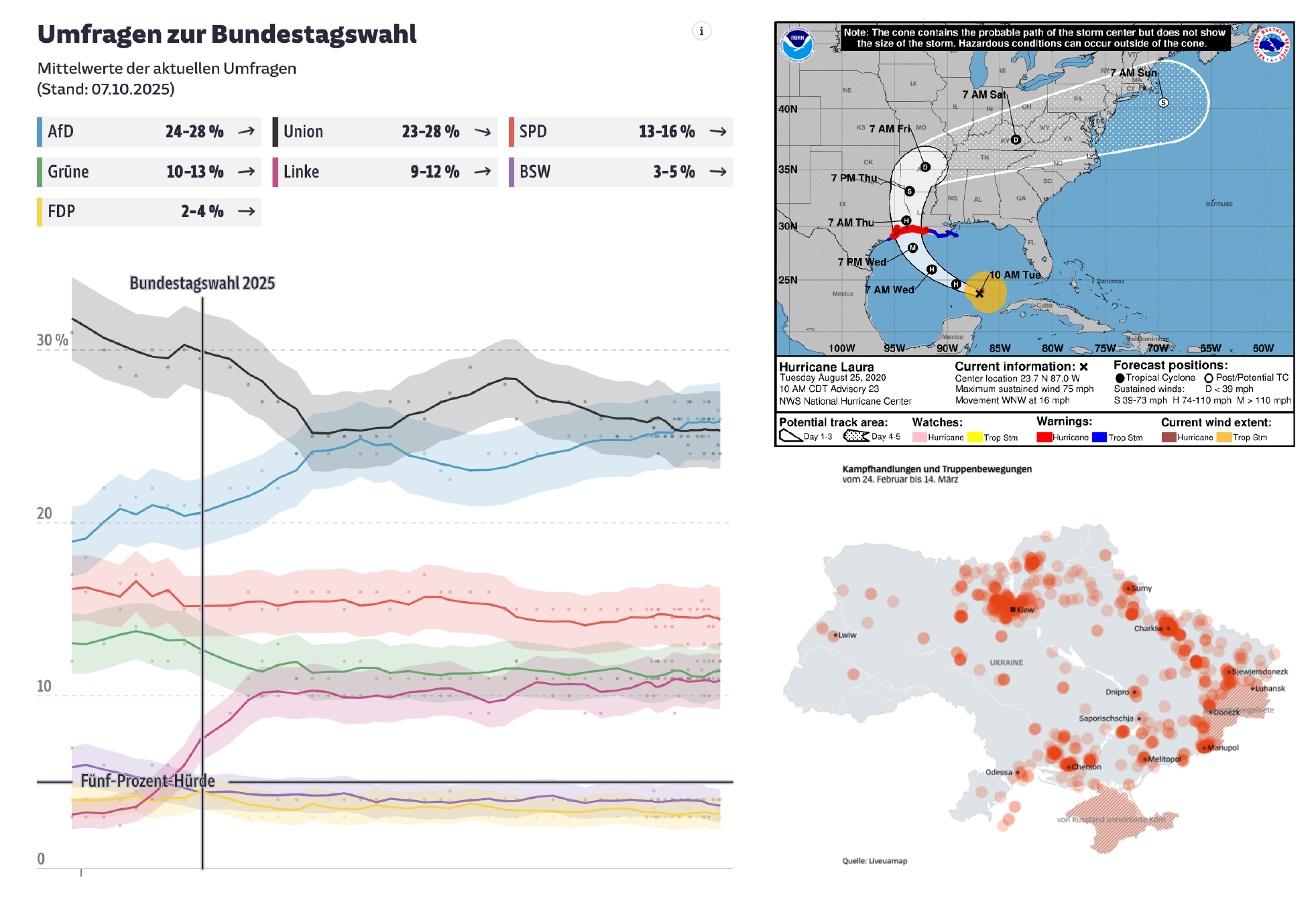}
    \caption{Application examples for uncertainty visualization most discussed by workshop participants (Phase \roundnum{color_phase1}{1}): election poll visualization
    \href{https://www.sueddeutsche.de/politik/bundestagswahl-regierungsbildung-umfragen-kleine-parteien-li.3202003?reduced=true}{[SZ]} (left), hurricane forecast maps \href{https://www.nhc.noaa.gov/aboutcone.shtml}{[NOAA]} (upper right), and conflict maps \href{https://www.spiegel.de/ausland/ukraine-krieg-der-aufmarsch-die-wende-die-aktuellen-gefechte-100-tage-bilanz-in-karten-a-3339361a-b4f8-4582-a768-c20ec47fdc4f}{[Der Spiegel]} (lower right).}
    \label{fig:commonexamples}
\end{figure}

\subsection[(F1) Newsroom Examples]
{\roundnum{color_phase1}{F1} Newsroom Examples }

During our discussions, participants shared insights about the practical use of uncertainty visualizations in their respective data journalism teams. 
The examples below illustrate how these considerations apply in participants' practice. 
Three recurring topics were discussed in most depth: election polls, hurricane path forecasts, and conflict zone mapping. These examples sparked detailed discussions around established practices, challenges, and anticipated audience interpretation.

\subsubsection{Election polls} 
Participants described election polls as a domain where uncertainty plays a central role but is difficult to communicate without substantially increasing visual complexity. They pointed out several sources of uncertainty such as sampling error, response biases, measurement effects, and weighting models. 

\textbf{Visualization approaches.} A recurring visualization strategy is to show a \textit{range instead of an exact percentage} (Figure~\ref{fig:commonexamples}). As one participant put it, \textit{``There is no single value to which the current polls can be pinned down.''} Another participant suggested \textit{``write the mean value first, followed by the range in brackets,''} to make comparisons easier. One participant mentioned using \textit{tooltips} to provide additional context without overloading the main visualization.

Other established methods include \textit{time series with trend lines} that have  \textit{``certainly become well established''} but may \textit{``suggest a level of accuracy that does not actually exist.''}  As an alternative, participants discussed plotting individual polls from different institutes as points only, hoping that \textit{``this will be enough [...] to convey the uncertainty.''} They also debated trend indicators such as arrows, raising questions such as \textit{``When should the arrow be shown, and in which direction? Does it make sense?''}. 
\textit{Simulation-based visualizations} were mentioned as well, for example to communicate majority probabilities, coalition likelihoods, or whether a party will exceed the 5\%-threshold, which is highly relevant in the German electoral system.

\textbf{Challenges.} Participants discussed several practical considerations \cg{that inform decisions about how} uncertainty is implemented in election polling visualizations. 
A recurring concern was visual density, for example, the visualization may become too dense when uncertainty bands are added to time series with multiple parties, especially for smaller parties. As one participant noted: \textit{``there is very, very much overlap and you already have difficulty distinguishing the lines in parts […] it simply becomes visually very, very full and complex if you additionally also display these uncertainty areas.''} For this reason, some editorial teams deliberately decide not to include them.

Editorial goals are also deemed to influence design decisions. In some cases, participants reported intentionally avoiding representations that could obscure the central message of the graphic. For example, one participant explained that their team chose not to emphasize overlapping ranges in the legend because they wanted to clearly communicate that one party was ahead of another: \textit{``we do not want to say […] the ranges overlap. We want to show: The Union [CDU/CSU, center-right party alliance in Germany] is ahead of the AfD [right-wing party in Germany].''}



\subsubsection{Hurricane paths} 
Participants described hurricane-path forecasts as a prominent case of uncertainty visualization, combining spatial and temporal prediction and appearing frequently in U.S. news coverage.

\textbf{Visualization approaches.} A classic representation is the \textit{cone of uncertainty}, which shows the most likely predicted path of a storm's center ~\cite{Ruginski02042016}  (Figure~\ref{fig:commonexamples}). However, as this representation tends to be misunderstood, participants discussed alternative formats that add additional forecast variables (e.g., the probability of hurricane-force winds in a region) and encode them with color gradients. Another participant mentioned ensemble-based approaches: \textit{``you can calculate the path of the eye of the storm by running 1{,}000 simulations of where it could pass through, and then displaying all these paths.''}

\textbf{Challenges.} Participants highlighted that uncertainty communication is closely tied to meteorological classification systems. For example, the distinction between tropical storms and hurricanes is based on predefined wind-speed thresholds. One participant noted that while these thresholds are precise, they do not necessarily convey broader context about the storm’s potential impact. This can make it challenging to communicate the practical meaning of the forecast through a single visual representation.

\subsubsection{Conflict zones} 
The third example involved uncertainty visualization in reporting combat operations within war zones. Unlike forecasts, the objective is to communicate uncertainty about \textit{where} combat activities occurred. Participants attributed this uncertainty to missing data  and differences in how event severity is assessed and weighted. 

\textbf{Visualization approaches.} One participant described mapping reported combat activities as overlapping symbols and aggregating them to spatial heat maps that reveal visible clusters rather than exact counts (Figure~\ref{fig:commonexamples}): \textit{``hot spots, or larger areas where it was possible to conclude that: Yes, based on the individual data points, there was more activity here, than elsewhere.''} Participants also discussed adding a temporal dimension (e.g., using scrollytelling to show the change over time). In cases where territorial control is ambiguous, they mentioned hatched areas as a convention.

\textbf{Challenges.} A key risk in conflict mapping is false precision when reporting incomplete data and a \emph{completeness illusion}: readers may assume that mapped points represent a comprehensive record of events.
As one participant put it, the goal was to prevent readers from thinking \textit{``all these individual points represent all the battles that took place there.''} Participants therefore treated aggregation and relative encodings as a way to communicate uncertainty inherent in incomplete reporting while still enabling comparison of areas with more versus less activity.

\textbf{Key findings \roundnum{color_phase1}{F1}:} Across examples (polling, hurricane forecasts, and conflict reporting), participants described uncertainty visualization as a pragmatic trade-off rather than a guideline-driven design task. They favored representations that reduce predictable misreadings, keep graphics readable despite added complexity, and stay in line with the central message. The three cases also illustrate that uncertainty visualization appears most often in domains where models, projections, or incomplete reporting are central, where design choices must balance transparency against risks of clutter and false precision.

\subsection[(F2) Uncertainty Scenarios]
{\roundnum{color_phase2}{F2} Uncertainty Scenarios}
\label{subsec:scenarios}
In the second phase of the workshop, participants applied our UUS framework (Table \ref{tab:uncertainty-scenarios}) to their own newsroom examples, revealing both the utility and limits of scenario-based classification in journalistic practice.

\textbf{Ambiguity of discrete classification. }Participants encountered difficulties in classifying the visualizations into discrete categories when dealing with the proposed UUS scenarios within the context of their own work. The discussion showed that a single visualization often represented multiple usage scenarios and one journalist said that they ``\textit{had the feeling that there is still a missing category}'', but could not identify what it should be. Participants proposed new scenarios based on their examples, such as \textit{research method uncertainty}, \textit{visualization uncertainty}, and \textit{scientific uncertainty}, suggesting gaps at the boundaries of the proposed set.

\textbf{Data- and model-centered perspective. }A recurring discussion was that participants tended to approach the usage scenarios from a data- and model-centered perspective. Although the framework was intended to focus on the different journalistic contexts of presenting uncertainty, statistical and model-related uncertainties within the data collection process were frequently mentioned. This reflects the position of data teams as an interface between statistical-methodological work and content-driven storytelling: when reasoning about uncertainty, they often emphasize technical aspects and constraints that determine what can credibly be claimed in a story.


\textbf{Scenario salience and distinguishability.} Among the proposed scenarios, the \textit{Polling } scenario (later expanded to \textit{Sampling}), was by far the most relevant in the workshop. It was discussed extensively and supported by several newsroom examples. Compared to other domains, uncertainty visualizations in the context of political polls appeared to be the least controversial usage scenario as they were more clearly distinguishable. Other subjects, such as COVID-19 rapid tests, war reporting, and cancer risk studies, apparently showed a higher degree of ambiguity and participants found it more difficult to decide which scenario best reflected the relevant type of uncertainty and communication intent.

\textbf{From taxonomy to planning prompt.} 
Another participant emphasized the value of the framework as a conceptual tool in the planning phase of visualization work, rather than as a taxonomy for classifying existing visualizations. In that role, the scenario framework can support questions such as: ``\textit{Is there a UUS scenario that fits my use case and what can I take away for my own visualization and my story?}'' Another participant also  noted that dealing with the different examples and usage scenarios helped to broaden their perspective on uncertainty, shifting from a singular notion of uncertainty toward a richer understanding of the variety of uncertainty visualization approaches and their purposes in data journalism practice.

\textbf{Key findings \roundnum{color_phase2}{F2}: }Participants found it difficult to classify examples into discrete scenarios while Polling/Sampling emerged as the most salient and least ambiguous scenario. Overall, participants tended to reason about scenarios in data- and method-centered terms and ultimately reframed the scenario framework mainly as a planning prompt to support early design decisions rather than a strict taxonomy for labeling finished visualizations.

\subsection[(F3) Authoring Hurdles]
{\roundnum{color_phase3}{F3} Authoring Hurdles}
Participants described a range of editorial, practical, and conceptual challenges associated with creating and integrating uncertainty visualizations into journalistic practice. These challenges relate to decision-making processes, limitations in data and expertise, production constraints, and difficulties in communicating uncertainty to audiences.

\textbf{Data availability and uncertainty quantification. }Participants highlighted data-related challenges. In some cases, uncertainty cannot easily be quantified or visualized because the underlying data do not provide confidence intervals or other statistical measures of uncertainty. Datasets based on estimates or incomplete information may lack agreed uncertainty metrics, and participants also reported difficulties in deciding which measure to use (e.g., which confidence interval is appropriate). In other cases, uncertainty arises from limited sample reliability or incomplete datasets, requiring careful interpretation before communication.

\textbf{Decision-making without systematic evaluation. }This challenge concerns the decision-making process about whether or not to include uncertainty. Several participants reported that the decision to include or omit uncertainty visualizations is often intuitive rather than systematic, largely due to limited experience and a lack of empirical audience research. As one participant noted, such decisions are frequently made without established evaluation frameworks, meaning that reflection on the effectiveness of a visualization often occurs only retrospectively.

\textbf{Production overhead under deadline pressure. }Participants frequently mentioned time pressure and workload, especially in daily news production. Because uncertainty visualizations often require additional data processing, design work, and explanatory effort, they are more likely to be used in larger, in-depth projects than in routine daily updates. As one participant
described, producing such visualizations can involve \textit{``increased time expenditure''}, which conflicts with the demands of topicality and rapid publication.

\textbf{Organizational routines and coordination effort. }Participants noted that the importance of uncertainty is perceived differently within editorial teams, and that producing such visualizations may involve coordination between multiple stakeholders, including journalists, designers, and editors. Some participants also described institutional inertia, referring to established routines in editorial production, for example the tendency to follow existing conventions or the perception that \textit{``we have always done it this way''}.

\textbf{Format and media channel constraints. }Participants emphasized that visualization design must be adapted to different publication formats and distribution channels. For example, a large print graphic offers different possibilities than a visualization designed for mobile platforms or social media. As one participant explained, \textit{``a page-sized print graphic simply looks different and offers different possibilities than an Insta post''}. Print formats can constrain space and force static layouts, whereas digital formats allow interactivity and more design flexibility but often operate under limited attention and mobile-first consumption.

\textbf{Transparency-complexity trade-offs. }Participants frequently mentioned the tension between transparency and complexity. Participants noted that uncertainty visualizations are often considered important when transparency is essential, for example in scientific reporting or in topics such as elections, pandemic modeling, climate change, or conflict reporting. In such cases, uncertainty visualization is seen as a way to show \textit{``what can actually be said with the data.''} At the same time, participants report that uncertainty may be omitted if it adds little value, increases complexity, or risks overloading the graphic.

\textbf{Communication challenges for heterogeneous audiences. }Explaining uncertainty in a compact and accessible way is described as difficult, particularly when visualizations rely on technical terminology or statistical concepts. Participants noted that uncertainty visualizations can increase ambiguity, potentially weakening the perceived clarity or news value of a story. 
Another important remark made in the workshop is that uncertainty often requires accompanying explanation in accessible language. Without sufficient textual or narrative explanation, readers may struggle to interpret visualizations correctly, particularly when statistical concepts are involved.
Some of the journalists also cautioned that visual representations of uncertainty---such as certain animations or overly complex simulations---can be misleading or cause readers to lose visual orientation if they are not carefully designed.


\textbf{Key findings \roundnum{color_phase3}{F3}:} Participants described authoring uncertainty visualizations as additional high-effort work: decisions are often made intuitively without established evaluation practices, uncertainty information and metrics are frequently unclear or unavailable, and tight production timelines, coordination effort, and editorial constraints limit what can be implemented. As a result, uncertainty is most likely to be included when it materially improves transparency, but is often simplified or omitted when it increases visual complexity or risks confusing audiences without sufficient explanatory support.

\subsection[(F4) Audiences]
{\roundnum{color_phase4}{F4} Audiences}
Participants emphasized that uncertainty visualizations in journalism are designed for a broad general-news readership with varying knowledge, interest, and trust. Accordingly, they discussed what audiences are expected to take away, where misunderstandings arise, and which design measures help keep uncertainty communication understandable without overloading the story.

\textbf{Intended Audience of Uncertainty Visualizations. }
Participants described the target audience as broad and heterogeneous, reflecting the readership of general news outlets. Visualizations are designed for readers with \textit{``some schooling, no more''}, rather than for specialized or expert audiences.
Within the readership, participants identified different levels of \textit{knowledge} and \textit{interest}. The key audience groups were seen in non-specialist readers without statistical training, and more knowledgeable or highly interested readers, who may engage more deeply with the data and expect additional informational value.
Some participants also referred to trust-sensitive or skeptical readers, for whom uncertainty may be interpreted as a lack of reliability rather than transparency.

\textbf{Audience Expectations and Intended Takeaways. }
Participants had limited insight into how audiences interpret uncertainty visualizations and therefore made few assumptions about readers' levels of statistical literacy.
As a baseline, they aimed for readers to understand the main message of the visualization, even if not all readers fully grasp the statistical concepts involved.
In election polling, for example, participants expected readers to recognize relative developments and plausible outcomes (e.g., parties gaining or losing support, potential seat distributions, coalition scenarios).
It was also stated that uncertainty visualizations should encourage more nuanced, non-deterministic interpretations by challenging binary readings and making variation and limits visible.
Participants emphasized that audience engagement and comprehension vary strongly depending on attention span, publication channel, and topic interest.

\textbf{Intended Benefits of Uncertainty Visualization for Readers. }
Participants identified several intended benefits of uncertainty visualizations for readers.
A frequently mentioned benefit was representing results more realistically, for example by showing ranges of possible outcomes rather than a single deterministic value.
Participants also viewed uncertainty visualizations as a way to make data collection, estimation, and modeling processes more transparent by highlighting assumptions and variability.
Participants also suggested that uncertainty visualizations can provide informational value across different levels of reader background. They described this as supporting a basic understanding for general audiences while still enabling deeper inspection for interested readers.

\textbf{Risks and Drawbacks when Confronting Audiences with Uncertainty Visualization. }
A central concern was misunderstanding or misinterpretation. Participants noted that uncertainty visualizations often require additional cognitive effort, and readers may struggle to interpret them correctly or may focus on individual elements rather than the overall message.
They also mentioned the possibility of increased cognitive complexity that may reduce reader engagement, for example causing readers to disengage or leave the page, with the ultimate risk of losing readers.
In politically sensitive contexts such as election polling, participants raised concerns about cherry-picking or deliberate misreadings.
Several participants further cautioned that communicating uncertainty may reduce perceived credibility or trust if audiences interpret uncertainty as unreliable information rather than as transparency.

\textbf{Measures and Design Strategies to Support Audience Understanding. }
\cg{Strategies discussed to support audience understanding include} the use of engaging and visually appealing designs that capture readers’ attention and encourage them to engage with the visualization.
Narrative techniques were also mentioned as a way to guide readers through complex information, e.g., \textit{scrollytelling}, which reveals the uncertainty step-by-step, explaining it as readers progress through the article. The importance of combining explanatory text or simplified language was also highlighted, as it supports interpretation, especially for readers without statistical expertise.
Simplified or easy-to-read language offers alternative text versions that present the same content in simpler terms.
Participants suggested minimizing visual complexity, avoiding \textit{``chartjunk''}, to keep the central message clear.
They emphasized that appropriate complexity depends on publication context: fast-paced platforms such as social media tend to require simpler formats, whereas recurring features (such as poll trackers) allow readers to become familiar with more complex representations over time.
Participants also noted that evaluating audience understanding remains difficult in practice. Journalists therefore often rely on indirect indicators such as click rates, completion rates, or reader comments to assess how well a visualization is received.

\textbf{Key findings \roundnum{color_phase4}{F4}: }
Participants described uncertainty visualizations as targeting a broad general-news readership with varying literacy, interest, and trust. While having limited insight into who engages with a visualization, they primarily aim for readers to grasp the central message and that outcomes are uncertain, while recognizing that engagement depends strongly on channel and attention. Participants saw benefits in representing results as ranges and making assumptions and variability visible, but worried about misinterpretation, cognitive overload, disengagement, cherry-picking in sensitive contexts, and potential trust loss. To mitigate these risks, participants emphasized combining uncertainty visualizations with concise explanatory text, using simplified language, and employing layered presentation techniques such as scrollytelling while reducing unnecessary visual complexity.

\subsection[(F5) Takeaways \& Future Directions]
{\roundnum{color_phase5}{F5} Takeaways \& Future Directions}

Participants described several areas in which further development of uncertainty visualization in journalism is considered necessary, including visualization approaches, interactive explanations, empirical evidence on audience understanding, and guidance for design.

\textbf{Application-specific uncertainty visualization approaches. }
Discussions around election poll visualizations highlighted ongoing experimentation and the lack of established approaches. While many teams have explored different ways of representing uncertainty, particularly during recent election coverage, participants noted that no widely accepted standard has emerged. As one participant stated, although many teams have ``found a good solution,'' none has produced a format that could be considered the definitive model for future coverage. As a result, they stated the need for establishing approaches for certain recurring applications such as poll reporting.


\textbf{Layered interaction and avoiding false precision. }
Participants highlighted interactive formats for communicating uncertainty as a promising direction to balance accessibility and complexity. Layered interaction was described as a way to provide a simple overview at first glance while enabling interested readers to explore additional details on demand.
In addition to interaction, participants discussed representational strategies that reduce the risk of false precision in high-uncertainty domains: for example, in conflict reporting, relative or qualitative summaries (e.g., increases/decreases) can be preferable to precise counts when the underlying data are incomplete.

\textbf{Understanding the audience better. }
Another recurring theme concerned the lack of empirical evidence on how audiences interpret uncertainty visualizations. Participants called for studies that compare visualization formats with and without uncertainty representations in order to assess differences in perceived understanding and information uptake.

\textbf{Improving design guidance. }
Finally, participants noted the absence of established best practices or guidelines for uncertainty visualization in journalism. Several expected that existing research might already provide clearer recommendations, e.g., in the form of documented best practices, and viewed the development of such practice-oriented guidance as an important direction for future work.


\textbf{Key findings \roundnum{color_phase5}{F5}: }Participants called for continued development of uncertainty visualization in journalism through ongoing experimentation and stronger evidence on audience interpretation. They highlighted election polling as a key area where no widely accepted standard has emerged for uncertainty visualization. Interactive, layered approaches were seen as promising for balancing accessibility and complexity, while in domains like conflict reporting participants suggested using relative or qualitative measures to avoid false precision. Overall, they emphasized the need for empirical studies comparing formats (with vs. without uncertainty) and for clearer, practice-oriented guidelines.

\section{Discussion}

Our findings align with prior work that uncertainty is frequently minimized or omitted in practice due to concerns about cognitive load, credibility, and the effort required to compute and explain uncertainty (e.g., \cite{Hullman:2019}). They also echo documented misinterpretations of common encodings, such as interval misunderstandings and the cone-of-uncertainty problem (e.g., \cite{Broad:2007,Ruginski02042016}), which participants cited as reasons to simplify representations or to pair visuals with explanation. 

Our workshop adds detail on how these issues are negotiated under newsroom conditions: \cg{uncertainty decisions are shaped not only by representational considerations but also by editorial intent, production overhead, and channel constraints.} 
Participants’ responses also highlight how existing frameworks are appropriated in practice: often as decision support rather than classification.


\subsection{General Insights}
Considering all findings, uncertainty visualization in data journalism appears less as a question of finding a single ``best'' encoding and more as a \textit{workflow- and context-dependent decision} under practical constraints, here, editorial intent (preserving a clear takeaway), anticipated audience interpretation (misreading, disengagement, trust loss), and production feasibility (time, coordination, media channel requirements). This helps explain why uncertainty is often communicated selectively and why solutions that are attractive in principle may be impractical in routine reporting.

\textbf{Pragmatic defaults dominate in practice.}
Participants described using low-overhead cues (e.g., ranges, aggregation, qualitative or relative summaries) and omitting more explicit representations when these increase clutter, require extensive explanation, or weaken narrative clarity (\roundnum{color_phase1}{F1}, \roundnum{color_phase3}{F3}).
This reveals that simplification is not merely a lack of standards but a rational adaptation to production realities: a trade-off between expressive precision and communicational safety under time pressure, channel limitations, and the need for clear takeaways. Rather than developing more sophisticated encodings in isolation, the focus may shift to tools and templates that embed uncertainty communication into actionable decisions, aligning with broader critiques regarding the limited usability of visualization research in real-world contexts~\cite{Kosara:2016}.

\cg{Prior literature highlights that data workers manage uncertainty throughout the general workflow rather than treating it as a purely representational challenge\cite{Boukhelifa2017}. Similarly, journalists in our study described uncertainty communication as a pragmatic editorial decision shaped by workflow constraints, audience considerations, and production overhead.}

\textbf{Misinterpretation risk functions as a design constraint.}
Participants emphasized that uncertainty encodings are not only challenging to design but also easy to misread without guidance (\roundnum{color_phase1}{F1}, \roundnum{color_phase4}{F4}). Concerns about false precision, over-reading of trend lines, or misunderstandings of forecast representations led them to prefer simpler encodings, paired with explanatory text or narrative guidance that helps readers interpret the uncertainty representation correctly (\roundnum{color_phase1}{F1}, \roundnum{color_phase4}{F4}). This reframes misinterpretation not as a problem to be solved by better encodings alone, but as a constraint that practitioners internalize early in design. While prior work shows that interpreting uncertainty requires additional cognitive effort~\cite{Allen:2014}, our findings suggest that practitioners operationalize this insight by treating explanation as integral to visualization, not as an add-on.

\cg{This perspective is consistent with Padilla et al., who found that people integrate both visual representations of quantitative uncertainty and qualitative expressions of confidence when making decisions, suggesting that explanatory framing contributes meaningfully to uncertainty interpretation rather than merely supplementing the visualization. \cite{padilla2021uncertain}}

\textbf{Audience heterogeneity is managed via layered communication strategies.}

Participants' accounts suggest that newsroom uncertainty communication is shaped by an implicit awareness of wide variation in statistical literacy, motivation, and institutional trust (\roundnum{color_phase4}{F4}). Strategies are intended for inclusion of diverse levels of statistical knowledge and understanding paces, alongside channel-adaptive simplification for social media and mobile contexts (\roundnum{color_phase3}{F3}, \roundnum{color_phase4}{F4}, \roundnum{color_phase5}{F5}). 

Rather than treating audience heterogeneity as a segmentation problem, practitioners tend to operationalize it through layered communication that embeds differentiation within a single artifact. While consistent with user-centered calls to account for differences in uncertainty interpretation~\cite{Bryan:2021,Dries:2025}, our findings show how such heterogeneity is negotiated under newsroom constraints of speed, attention scarcity, and platform logic. These strategies do not merely illustrate the HCI principle of progressive disclosure~\cite{Springer:2020}, but how it is repurposed as an editorial compromise between transparency and narrative clarity. Layering becomes not only a usability technique, but a mechanism for balancing epistemic responsibility with communicative efficiency.

\cg{The layered communication strategies described by participants also follow recent calls for greater user agency in uncertainty communication. Peña-Araya et al. argue that uncertainty should not always be treated as a fixed property to be conveyed, but as something users actively interpret and negotiate. From this perspective, newsroom techniques such as progressive disclosure through scrollytelling and optional interactive details can be understood as mechanisms for supporting different levels of uncertainty engagement among heterogeneous audiences. \cite{Penya2025}}

\textbf{Frameworks matter when they support early design decisions.}
Participants' call for application-specific guidance does not contradict existing research frameworks. It reveals a gap between analytic models and operational support needed in newsrooms (\roundnum{color_phase2}{F2}, \roundnum{color_phase5}{F5}). While frameworks help structure communication and thinking about uncertainty visualization, practitioners often need actionable defaults that translate principles into feasible choices under constraints of practical work. 
Our findings suggest that the UUS framework (Table \ref{tab:uncertainty-scenarios}) can help close this gap when treated as a \textit{planning prompt}. Participants were critical of using it for discrete scenario classification but they valued the framework for structuring early design reasoning and stimulating discussion and critical reflection on how uncertainty should be communicated (\roundnum{color_phase2}{F2}). 

Based on this, we suggest guidance at the \textit{scenario level} rather than for individual applications. Scenario-level guidelines (e.g., \textit{Sampling} instead of \textit{Election polls}) offer a useful common ground: general enough to transfer across cases, yet specific enough to inform practical design choices. 
This reinforces prior observations of a gap between visualization research and practice, where the challenge is less a lack of awareness than the presence of competing constraints that limit applicability in real-world settings~\cite{Correll:2022}. In such contexts, frameworks become most useful when they support early-stage reasoning rather than prescribing rigid classifications, suggesting a need for approaches that translate abstract principles into flexible, context-sensitive guidance. 

\section{Two-level design guidance}
\cg{We provide two-level design guidance for uncertainty visualization in data journalism consisting of the UUS framework and design heuristics. UUS supports scenario-level reasoning about what kind of uncertainty is central to a story, while the heuristics support concrete editorial and design decisions about whether and how to communicate that uncertainty.}

\cg{Building on our synthesized findings, we derive a set of practice-oriented heuristics (Table \ref{tab:heuristics}). The heuristics are grounded in the workshop evidence (F1–F5) and are organized into three categories: Before, during, and after the design process. They are formulated as rules-of-thumb to support context-sensitive design decisions and serve as a basis for future, more detailed practice-oriented guidelines.}

\subsection{Revised UUS Framework}
\cg{Building on Phase~\roundnum{color_phase2}{2}, we revised the Uncertainty Usage Scenarios (UUS) framework based on participants' discussions after applying it to their own examples. The refinement primarily addressed three aspects. First, several examples could meaningfully be associated with more than one scenario. We therefore treat the scenarios as non-mutually exclusive rather than as discrete categories. Second, we revised terminology in response to the discussions, including renaming \textit{Changing Conditions} to \textit{Definite Ranges} and \textit{Polling} to \textit{Sampling}. Third, we consolidated rarely used or overlapping categories by merging the initial \textit{Reconstructed Past} category into \textit{Missing Data}.}

\cg{Table~\ref{tab:uncertainty-scenarios} reports the refined set with definitions and newsroom examples discussed during the workshop. Based on participants' difficulties with discrete classification and their reflections on how the framework could support practice, we frame UUS as a planning-oriented structure for early design decisions rather than as a strict taxonomy for labeling finished visualizations. In this role, UUS can support reasoning about which uncertainty is relevant to a story, what aspects of that uncertainty need to be communicated, and how it might be represented under editorial and publication constraints.}

\subsection{Practice-oriented Heuristics}


From our synthesized findings we derive a set of \textit{practice-oriented heuristics} (Table \ref{tab:heuristics}), grounded in the workshop evidence (\roundnum{color_phase1}{F1} to \roundnum{color_phase5}{F5}), reflecting participants' discussions in relation to the current state of research. \cg{In the findings, we looked for central insights and suggestions that can be integrated into practical workflows.} 

\cg{We organize the heuristics according to three stages of the design process: \textit{before design}, addressing whether and what uncertainty should be communicated; \textit{during design}, addressing how uncertainty is represented and explained for different audiences and publication contexts; and \textit{after design}, addressing the evaluation of audience understanding.}

\cg{In addition to the description in this paper, we created a \textit{cheat sheet} for practitioners to be able to remember and use the heuristics without the grounding and justification in Table 2.} 

\cg{The heuristics} should be understood as rules-of-thumb rather than universally applicable rules. \cg{In particular, they do not prescribe specific visual encodings for particular types of uncertainty. Instead, they capture important decision points and trade-offs that emerged across the workshop, such as whether uncertainty affects the story, how additional complexity can be managed, and how audience interpretation can be supported. For each heuristic, Table~\ref{tab:heuristics} describes its empirical basis, distinguishing between practices participants reported already applying and insights they expressed based on professional experience or intuition. It also provides theoretical grounding by relating each heuristic to prior research.}

\cg{We therefore view the heuristics as practical decision support as well as a starting point for further validation. They can support reflection and discussion in newsroom production, while heuristics currently grounded primarily in practitioner experience or intuition identify opportunities for future empirical evaluation.} To our knowledge, this is one of the first practitioner-derived heuristic sets explicitly tailored to uncertainty visualization decisions under newsroom production constraints.

\section{Limitations and Future Work}
Our findings are grounded in a single workshop with a limited set of participants and are therefore not generalizable. Our results reflect the perspectives of participants from German-speaking newsrooms and may not transfer directly to other domains or organizational settings. \cg{We did not systematically collect participant-level characteristics such as years of professional experience, specific areas of specialization, or standardized measures of familiarity with visualization and uncertainty. We therefore cannot assess how such differences may have affected participants’ perspectives or the workshop findings.} Still, we see clear value in this bottom-up, practitioner-centered approach---the workshop surfaced how uncertainty visualization decisions are actually made under newsroom conditions. 

As with any workshop study, group dynamics (e.g., social desirability) may have determined what was discussed, and our data capture only part of day-to-day practice. Still, direct engagement with practitioners provides access to tacit knowledge, concrete workarounds, and domain-specific requirements that are difficult to capture otherwise. 

Future work should validate and refine our heuristics across additional newsroom settings, and through audience studies comparing alternative encodings and explanatory strategies of uncertainty visualization in realistic journalistic settings, including heterogeneous audience groups. In addition, comparative work on uncertainty communication across domains (e.g., public health and climate science) could help identify which challenges and solutions generalize and which remain context-specific. 
Building on our finding that participants framed the UUS framework as a planning prompt, future studies should evaluate whether using UUS or similar frameworks during production improves design choices within newsroom workflows. 
Finally, developing newsroom-feasible evaluation methods beyond engagement metrics would help practitioners assess interpretive outcomes and iteratively improve uncertainty visualization in practice.

\section{Conclusion}
Uncertainty visualization research often advances new visualization methods and frameworks, yet adoption in day-to-day practice remains limited. Our workshop findings shed light on this gap by showing that uncertainty communication in data journalism is primarily a \emph{production decision} negotiated under constraints: editorial intent, limited time and coordination capacity, channel-specific formats, and the risk of predictable audience misreadings. Under these conditions, practitioners rarely aim for maximum visual or statistical expressiveness. Instead, they choose visualization formats that are communicatively ``safe'' (easy to read, explainable, and consistent with the story) and simplify or omit uncertainty when it would introduce clutter, weaken the takeaway message, or require explanation they cannot reliably provide.

This perspective also reframes the role of research frameworks in practice: our results indicate that frameworks like our UUS are difficult to apply as strict taxonomies, but valuable as planning prompts that structure early design reasoning and help teams identify relevant uncertainty for a story. At the same time, practitioners request guidance for specific applications. Building on this, we suggest developing guidance at the scenario level as practical middle ground between abstract frameworks and highly specific, case-by-case advice.

Finally, we contribute a set of practice-oriented heuristics derived from practitioner discussions. Instead of claiming universal validity, these heuristics operationalize recurring trade-offs and decision criteria reported by participants and can serve as a shared vocabulary between visualization research and newsroom practice. More broadly, our work suggests that uncertainty visualization research should also be assessed with respect to workflow fit and interpretive outcomes in realistic practical settings, motivating future work on reusable templates and practice-feasible evaluation methods.

\section{Acknowledgments}
This work was supported as part of project \textit{uncertainty4ddj} funded by the German Federal Ministry of Research, Technology and Space (BMFTR), grant number 03DPS1120.

\cg{The authors used ChatGPT (OpenAI) to assist with language editing, proofreading, and improving the clarity of the manuscript. it was also used to generate the images in Table 1. All AI-assisted revisions were reviewed and approved by the authors, who take full responsibility for the final content.}


\label{sec:references_inst}
\bibliographystyle{abbrv-doi-hyperref}

\bibliography{uncertainty-workshop}

\begin{table*}[t]
\centering
\small
\renewcommand{\arraystretch}{0.95}
\setlength{\tabcolsep}{4pt}

\begin{tabularx}{\textwidth}{
>{\raggedright\arraybackslash}m{1.3cm}
>{\hsize=1\hsize\raggedright\arraybackslash}X
>{\hsize=1\hsize\raggedright\arraybackslash}X
>{\hsize=1\hsize\raggedright\arraybackslash}X
}
\toprule
\textbf{Scenario} &
\textbf{Schema} &
\textbf{Definition} &
\textbf{Example from Workshop} \\
\midrule

\textbf{Forecast} &
\includegraphics[width=0.82\linewidth, height=3.0cm, keepaspectratio]{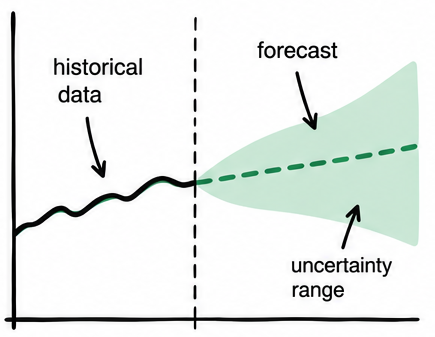} &
Showing that statements rely on assumptions about an unknown future. Uncertainty about future outcomes, based on predictions from models that depend on assumptions and input data. &
An interactive calculator that shows how long it would take the reader to become a millionaire.
\cg{A black line shows projected wealth over time, uncertainty is represented by a gray area around the line.}
\newline\cg{\textit{Source: \href{https://www.faz.net/aktuell/finanzen/millionaer-rechner-so-erhoehen-sie-ihr-vermoegen-auf-eine-million-euro-110310349.html}
{faz.de, ``Millionärsrechner -- So werden Sie Millionär'', Feb. 2025}}} \\

\midrule

\textbf{Definite Ranges} &
\includegraphics[width=0.82\linewidth, height=3.0cm, keepaspectratio]{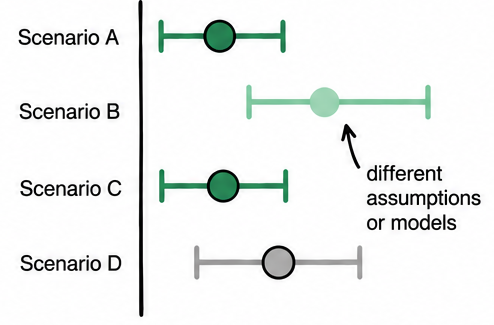} &
Estimating the expected reality within a range of possible outcomes. Uncertainty introduced when different assumptions, parameter settings, or external conditions change results.
\newline\cg{\sout{Originally named ``Changing Conditions'', renamed after the workshop.}} &
Projected changes of global temperature resulting from different political agreements \cg{(also classified as \textit{Forecast}).} 
\cg{Uncertainty is represented by markers indicating multiple projected temperature outcomes associated with different policy assumptions.
\newline\textit{Source: \href{https://www.sueddeutsche.de/projekte/artikel/wissen/klimawandel-konferenz-baku-cop29-forscher-warnung-klimazukunft-e140516/}
{sueddeutsche.de, ``Die Uhr tickt'', Nov. 2024}}} \\

\midrule

\textbf{Spatial} &
\includegraphics[width=0.82\linewidth, height=3.0cm, keepaspectratio]{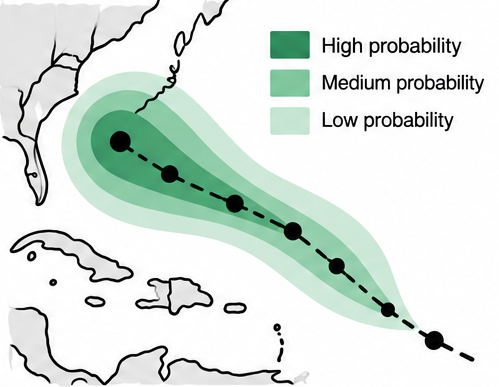} &
Estimating indeterminate locations, areas, etc. Uncertainty related to geographic data such as incomplete measurements, interpolation errors, or fuzzy boundaries. &
Projected path of a hurricane \cg{(also classified as \textit{Forecast}).}
\cg{Uncertainty represented by spatial probability regions around a projected path, with color intensity signifying probability level.
\newline\textit{Source: \href{https://www.nzz.ch/panorama/warum-die-gaengigste-hurrikan-visualisierung-problematisch-ist-und-was-die-nzz-an-ihrer-stelle-macht-ld.1394033}
{nzz.ch, ``Warum die gängigste Hurrikan-Visualisierung problematisch ist -- und was die NZZ an ihrer Stelle macht'', Nov. 2018}}} \\

\midrule

\textbf{Sampling} &
\includegraphics[width=0.82\linewidth, height=3.0cm, keepaspectratio]{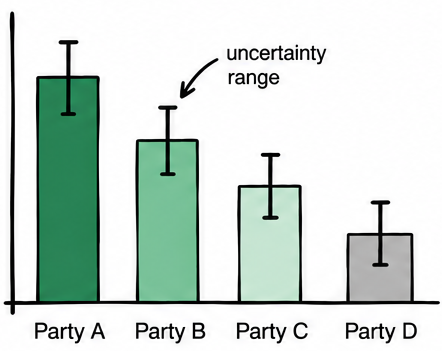} &
Assessing the results of a sample group and possible differences from reality. Uncertainty arising from surveys and polls including sampling error, response bias, or methodological flaws.
\newline\cg{\sout{Originally named ``Polling'', renamed after the workshop.}} &
Election poll results for the German federal election. \cg{Uncertainty is represented by hatched regions extending from bar estimates, indicating likely ranges.
\newline\textit{Source: \href{https://www.spiegel.de/politik/deutschland/sonntagsfrage-umfragen-zu-bundestagswahl-landtagswahl-europawahl-a-944816.html}
{spiegel.de, ``SPIEGEL-Umfrage-Tracker'', Feb. 2025}}} \\

\midrule

\textbf{Missing Data} &
\includegraphics[width=0.82\linewidth, height=3.0cm, keepaspectratio]{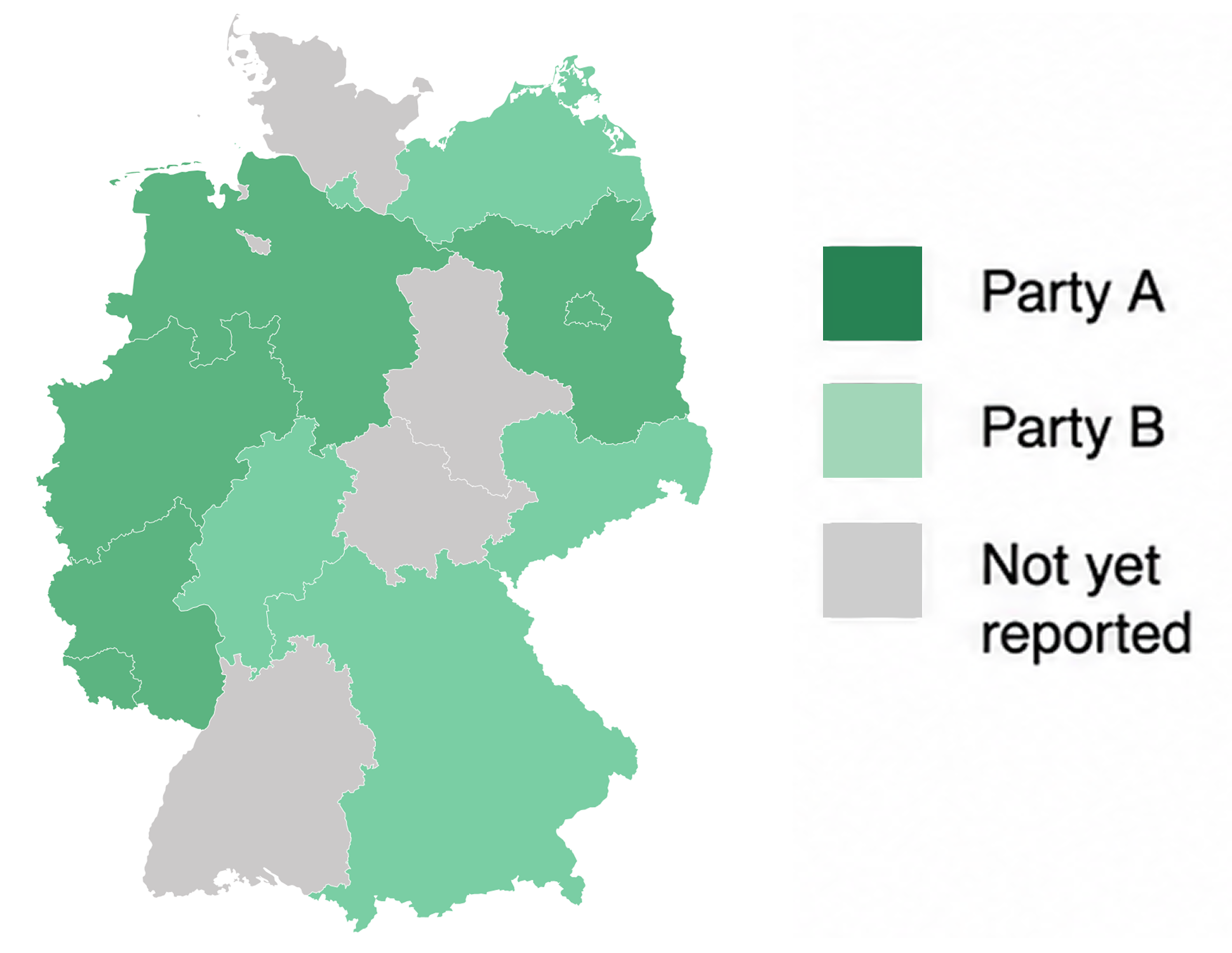} &
Assessing the limitations of a statement because certain information is not available. Uncertainty caused by absent, incomplete, or selectively reported data.
\newline\cg{\sout{Merged with another initial scenario (``Reconstructed Past'') after the workshop.}} &
Incomplete US election results reported up to the time of publication. 
\cg{Uncertainty is represented by gray-filled states with colored outlines, indicating states for which the outcome had not yet been decided.
\textit{\newline Source: spiegel.de, ``USA-Wahlen 2024'', Nov. 2024, not available online.}} \\

\midrule

\textbf{Classification} &
\includegraphics[width=0.82\linewidth, height=3.0cm, keepaspectratio]{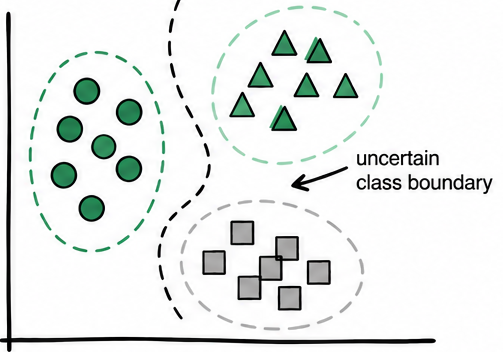} &
Uncertainty in assigning data to discrete categories, including ambiguous or borderline cases. &
\cg{The \textit{Classification} scenario did not occur in the examples discussed by participants. Example from the initial review: Self-positioning of EU election candidates along the dimensions of attitudes toward Russia and climate protection. Uncertainty is represented by contour lines.
\newline\textit{Source: \href{https://www.zeit.de/politik/deutschland/2024-06/europawahl-2024-kandidaten-einstellungen-russland-klimawandel}
{zeit.de, ``Klimaschutz und Russland: Wo stehen Sie?'', Jun. 2024}}} \\

\bottomrule
\end{tabularx}

\caption{Uncertainty Usage Scenarios (UUS) framework, \cg{including scenario definitions and newsroom examples discussed during the workshop.}}
\label{tab:uncertainty-scenarios}
\end{table*}


\clearpage
\onecolumn

\begin{landscape}

\small
\renewcommand{\arraystretch}{1.7} 

\begin{longtable}{
>{\bfseries\raggedright\arraybackslash}p{0.14\linewidth}
>{\raggedright\arraybackslash}p{0.20\linewidth}
>{\raggedright\arraybackslash}p{0.20\linewidth}
>{\raggedright\arraybackslash}p{0.20\linewidth}
>{\raggedright\arraybackslash}p{0.20\linewidth}
}

\caption{Practice-oriented heuristics for uncertainty communication in newsroom contexts, illustrated with example design decisions. The heuristics as an illustrated, compact version (``cheat sheet'') are available at \href{https://viulab-hh.github.io/u4ddj-heuristics/}{https://viulab-hh.github.io/u4ddj-heuristics/}.}
\label{tab:heuristics}
\\

\toprule
\textbf{Heuristic} &
\textbf{Description} &
\textbf{Example\textsuperscript{a}} &
\textbf{Support from Findings\textsuperscript{b}}&
\cg{\textbf{Literature Support}}
\\
\midrule
\endfirsthead

\toprule
\textbf{Heuristic} &
\textbf{Description} &
\textbf{Example\textsuperscript{a}} &
\textbf{Support from Findings\textsuperscript{b}} &
\cg{\textbf{Literature Support}}
\\
\midrule
\endhead

\midrule
\multicolumn{5}{r}{Continued on next page}
\\
\endfoot

\bottomrule
\endlastfoot


\multicolumn{5}{l}{\textbf{\textit{Before Design Process}}}
\\
\midrule

H1 -- Necessity Check
&
\cg{Evaluate whether and how uncertainty visualization affects the interpretation of the story. Include explicit visual uncertainty when it substantially affects the interpretation or takeaway. Otherwise, use a representation of uncertainty that adds less complexity.}
&
If the story is ``party A clearly leads'', \cg{the chart communicates uncertainty only through a textual annotation}. If the story is ``race is close, no reliable lead'', the chart displays ranges because uncertainty directly affects the interpretation.
&
\roundnum{color_phase1}{F1} \cg{\circnum{black}{P}: Participants see uncertainty visualization as a pragmatic trade-off and intentionally omit uncertainty when graphics become visually dense.}

\roundnum{color_phase3}{F3} \cg{\circnum{black}{E}: Decisions to include uncertainty are often made without a systematic evaluation, balancing transparency against complexity.}
&

\cg{\textbullet\ Visualization authors often avoid uncertainty because of concerns about confusing viewers or adding complexity to an already information-dense display\cite{Hullman:2019}.}


\cg{\textbullet\ Professionals make context-dependent decisions on whether uncertainty is sufficiently relevant to influence analysis and decision-making\cite{Boukhelifa2017}}
\\

\midrule

\cg{H2 -- Consistent Uncertainty Formats}
&
Reuse consistent uncertainty conventions across recurring or related visualizations when they represent the same uncertainty construct. Consistency can help readers build familiarity with how uncertainty is encoded over time.
&
\cg{
Readers encounter the same uncertainty-range format across recurring election-poll coverage, such as a poll tracker and election articles published throughout a campaign. Because the uncertainty encoding remains consistent, readers can build familiarity with how polling uncertainty is represented.}
&
\roundnum{color_phase4}{F4} \cg{\circnum{black}{E}: Participants believe that recurring uncertainty visualizations can help readers become familiar with them.}

\roundnum{color_phase5}{F5} \cg{\circnum{black}{E}: Participants call for established approaches to visualize uncertainty.}
&

\cg{\textbullet\ Graph familiarity is one factor associated with how audiences understand visualizations\cite{Saske:2026}}
\\

\midrule

\cg{H3 -- Uncertainty Source Attribution}
&
Explicitly identify the source of uncertainty to support accurate reader interpretation. Clarify whether it stems from sampling error, model assumptions, or incomplete data, and translate technical causes into accessible language.
&
\cg{A short explanatory note states that the range reflects sampling uncertainty, while differences between polling institutes may also stem from methodological choices.}
&
\roundnum{color_phase1}{F1} \cg{\circnum{black}{P}: Participants distinguish between different sources of uncertainty (including sampling, weighting and response bias) when choosing visual representations.}

\roundnum{color_phase2}{F2} \cg{\circnum{black}{E}: Participants reason about uncertainty from data- and model-centered perspectives.}
&
\cg{\textbullet\ Uncertainty communication should account for the source of uncertainty as a distinct dimension of what is communicated\cite{vanDerBles2019}} 

\cg{\textbullet\ Data workers distinguish between sources and types of uncertainty, which influence how they manage uncertainty during analysis\cite{Boukhelifa2017}}
\\


\multicolumn{5}{l}{\textbf{\textit{During Design Process}}}
\\
\midrule

\cg{H4 -- Uncertainty-Framed Insights}
&
\cg{Design the uncertainty visualization in relation to the intended primary insight. Use visual hierarchy and accompanying text to anchor interpretation and avoid obscuring the intended message.}
&
\cg{The headline communicates the primary takeaway (``Party A remains ahead of Party B''), while the uncertainty range is shown as supporting context rather than competing with the main message.}
&
\cg{\roundnum{color_phase1}{F1} \cg{\circnum{black}{P}: Participants adapt uncertainty representations to preserve the intended message of the visualization.}}

&
\cg{\textbullet\ Narrative visualization uses visual emphasis and framing to guide attention toward intended messages\cite{SegelHeer:2010}}

\cg{\textbullet\ Rhetorical techniques such as framing, omission, aggregation, and annotations influence how readers interpret a visualization\cite{Diakopoulous_Hullman:2011}}
\\

\midrule

\cg{H5 -- Textual Framing}
&
\cg{Integrate uncertainty representations with explanatory text where needed rather than relying on visuals alone. Explanatory text can clarify what the uncertainty represents and reduce the risk of misinterpretation.}
&
\cg{The chart is accompanied by a short explanatory sentence stating that the shaded range indicates a range of plausible values rather than an exact result.}
&
\roundnum{color_phase4}{F4} \cg{\circnum{black}{P}: Participants use concise explanatory text and simplified language to reduce the risk of misinterpretation.}
&
\cg{\textbullet\ Explanatory annotations can influence readers' comprehension and interpretation of visualizations\cite{Rahman:2025}}


\cg{\textbullet\     People can integrate qualitative expressions of confidence with visualized quantitative uncertainty when making decisions\cite{padilla2021uncertain}}
\\

\midrule

H6 -- Progressive Disclosure
&
\cg{Structure uncertainty information so readers encounter it progressively rather than all at once. Layered presentation techniques, such as scrollytelling or optional interactive details, can manage information complexity while allowing interested readers to explore additional information.}
&
\cg{The initial chart presents the uncertainty range with a brief explanation. Additional details, such as sampling methods, polling institutes, or methodological assumptions, become available through scrollytelling or optional interaction.}
&

\roundnum{color_phase4}{F4} \cg{\circnum{black}{P}: Participants use scrollytelling and layered explanations to present complex information progressively.}

\roundnum{color_phase5}{F5} \cg{\circnum{black}{E}: Participants identify layered interaction as a promising way to provide additional detail on demand.}
&
\cg{\textbullet\ Narrative visualization combines author-driven guidance, sequencing and interactive exploration to help readers navigate information\cite{SegelHeer:2010}}

\cg{\textbullet\ The ``overview first, zoom and filter, then details-on-demand'' principle supports progressive access to complex information\cite{Schneiderman:1996}}

\cg{\textbullet\ User agency can allow readers to engage with uncertainty according to their information needs\cite{Penya2025}}
\\

\midrule

H7 -- Medium Adaptation
&
\cg{Adapt uncertainty representations to the constraints and affordances of each publication medium. Preserve interpretability when representations are resized, simplified, or stripped of interaction.}
&
\cg{The web version shows interactive multi-party trend lines with uncertainty bands and details on demand, whereas a social media version uses a simplified static chart with uncertainty ranges and short textual explanations.}
&

\roundnum{color_phase3}{F3} \cg{\circnum{black}{P}: Participants adapt visualization choices to publication formats such as web, mobile, and print.}

&
\cg{\textbullet\ Responsive visualization requires trade-offs between graphical density, readability, and preservation of the intended message across screen sizes\cite{Kim:2021}}
\\

\midrule

H8 -- Literacy Accessibility
&
Design for heterogeneous levels of audience numeracy and statistical literacy. Use visual clarity and plain language, minimize unnecessary jargon, and explain unavoidable technical concepts in accessible terms.
&
\cg{An explanatory note describes a 95\% confidence interval in plain language rather than relying on the technical term alone.}
&
\roundnum{color_phase4}{F4} \cg{\circnum{black}{E}: Participants consider differences in readers' statistical literacy an important factor in uncertainty communication.}

\roundnum{color_phase4}{F4} \cg{\circnum{black}{P}: Participants use plain language, avoid jargon, and simplify explanations for broader audiences.}
&
\cg{\textbullet\ Audience characteristics such as numeracy and familiarity with graphical formats are associated with visualization understanding~\cite{Saske:2026}.}

\cg{\textbullet\ Uncertainty representations can require adaptation to audiences with different expertise and information needs~\cite{Goodwin:2024}}.
\\

\midrule

H9 -- Trust Trade-Offs
&
\cg{Recognize that communicating uncertainty can strengthen trust for some audiences while weakening it for others. Frame uncertainty to clarify why it is being communicated and how confidently the information should be interpreted.}
&
\cg{The chart explains the role of uncertainty with statements such as ``Polls are estimates, not outcomes'' and ``Range reflects survey uncertainty''.}
&
\roundnum{color_phase4}{F4} \cg{\circnum{black}{E}: Participants express concern that communicating uncertainty may reduce readers' trust.}

\roundnum{color_phase4}{F4} \cg{\circnum{black}{E}: Participants consider potential effects on audience trust when deciding whether and how to communicate uncertainty.}
&

\cg{\textbullet\ Communicating uncertainty does not necessarily reduce trust and can preserve or increase perceived trustworthiness under some conditions\cite{vanderbles2020effects}.}

\cg{\textbullet\ Visualized uncertainty may decrease trust in data, particularly when uncertainty is high\cite{Srivastava:2026}}
\\

\midrule


\multicolumn{5}{l}{\textbf{\textit{After Design Process}}}
\\
\midrule

H10 -- Comprehension Evaluation
&
Evaluate whether readers interpret uncertainty as intended rather than relying solely on engagement metrics. Where feasible, complement behavioral metrics with measures that directly assess comprehension.
&
\cg{An in-article micro-check asking readers what an uncertainty range means could reveal misunderstandings that clicks and reading-completion metrics cannot.}
&
\roundnum{color_phase4}{F4} \cg{\circnum{black}{P}: Participants use clicks, completion and comments to measure engagement.}

\roundnum{color_phase5}{F5} \cg{\circnum{black}{E}: Participants call for empirical audience studies to assess how uncertainty visualizations are understood.}
&
\cg{\textbullet\ Visualization comprehension concerns the alignment between designers' communicative goals and viewers' interpretation\cite{Quadri:2024}}



\cg{\textbullet\ Visualization understanding requires evaluation across multiple dimensions rather than relying on a single performance measure\cite{Saske:2026}}
\\

\end{longtable}

\vspace{0.5em}

\footnotesize
\textsuperscript{a}
Examples refer to design decisions for a typical election poll chart, as shown in Figure \ref{fig:commonexamples}.

\textsuperscript{b}
Support from findings summarizes the empirical basis for each heuristic. 
References to findings (\roundnum{color_phase1}{F1}--\roundnum{color_phase5}{F5}) are accompanied by brief explanations of how each finding informed the heuristic. 
The evidence is differentiated into two forms of support:
\circnum{black}{P} for Current \textbf{P}ractice \textit{(participants reported applying this in practice)}, and \circnum{black}{E} for \textbf{E}xperience and/or Intuition \textit{(participants expressed this based on professional experience or intuition)}.

\end{landscape}

\twocolumn
\clearpage


\FloatBarrier

\newpage

\end{document}